# The Effect of Sunspot Weighting

Leif Svalgaard[1] (leif@leif.org), Marco Cagnotti[2], Sergio Cortesi[2]

[1] Stanford University, Cypress Hall C13, W.W. Hansen Experimental Physics Laboratory, Stanford University, Stanford, CA 94305, USA

[2] Specola Solare Ticinese, Via ai Monti 146, CH-6600 Locarno, Switzerland

**Abstract:**

Waldmeier in 1947 introduced a weighting (on a scale from 1 to 5) of the sunspot count made at Zurich and its auxiliary station Locarno, whereby larger spots were counted more than once. This counting method inflates the relative sunspot number over that which corresponds to the scale set by Wolfer and Brunner. Svalgaard re-counted some 60,000 sunspots on drawings from the reference station Locarno and determined that the number of sunspots reported were 'over counted' by 44% on average, leading to an inflation (measured by a weight factor) in excess of 1.2 for high solar activity. In a double-blind parallel counting by the Locarno observer Cagnotti, we determined that Svalgaard's count closely matches that of Cagnotti's, allowing us to determine the daily weight factor since 2003 (and sporadically before). We find that a simple empirical equation fits the observed weight factors well, and use that fit to estimate the weight factor for each month back to the introduction of weighting in 1947 and thus to be able to correct for the over-count and to reduce sunspot counting without weighting to the Wolfer method in use from 1893 onwards.

-----------------------------------------------------------------------------------------------------------

"Hoc opus, hic labor"

"When we wish to correct with advantage, and to show another that he errs, we must notice from what side he views the matter, for on that side it is usually true, and admit that truth to him, but reveal to him the side on which it is false. He is satisfied with that, for he sees that he was not mistaken, and that he only failed to see all sides. Now, no one is offended at not seeing everything; but one does not like to be mistaken, and that perhaps arises from the fact that man naturally cannot see everything, and that naturally he cannot err in the side he looks at, since the perceptions of our senses are always true."
Blaise Pascal, *Pensées* (1670).

## 1. Introduction

In 1945 Max Waldmeier became Director of the Zürich Observatory. In 1961, Waldmeier published the definitive Zürich sunspot numbers up until 1960 (Waldmeier 1961). He noted that "Wolf counted each spot – independent of its size – but single. Moreover, he did not consider very small spots, which are visible only if the seeing is good. In about 1882 Wolf's successors changed the counting method, which since then has been in use up to the present. This new method counts also the smallest spots, and those with a penumbra are weighted according to their size and the structure of the umbra". In 1968 Waldmeier (1968b, 1948) codified the weighting scheme as follows "Später wurden den Flecken entsprechend ihrer Größe Gewichte erteilt: Ein punktförmiger Fleck wird einfach gezählt, ein größerer, jedoch nicht mit Penumbra versehener Fleck erhält das statistiche Gewicht 2, ein kleiner Hoffleck 3, ein größerer 5".[1] However, Wolfer in 1907 (Wolfer, 1907) explicitly states: "Notiert ein Beobachter mit seinem Instrumente an irgend einem Tage $g$ Fleckengruppen mit insgesamt $f$ Einzelflecken, ohne Rücksicht auf deren Grösse, so ist die daraus abgeleitete Relativzahl jenes Tages $r = k(10g+f)$".[2] We can verify that Wolfer, contrary to Waldmeier's assertion that the Zürich observers began to use weighting "around 1882", did not weight the spots according to Waldmeier's scheme by comparing Wolfer's recorded count with sunspot drawings made elsewhere, e.g. Figure 1.

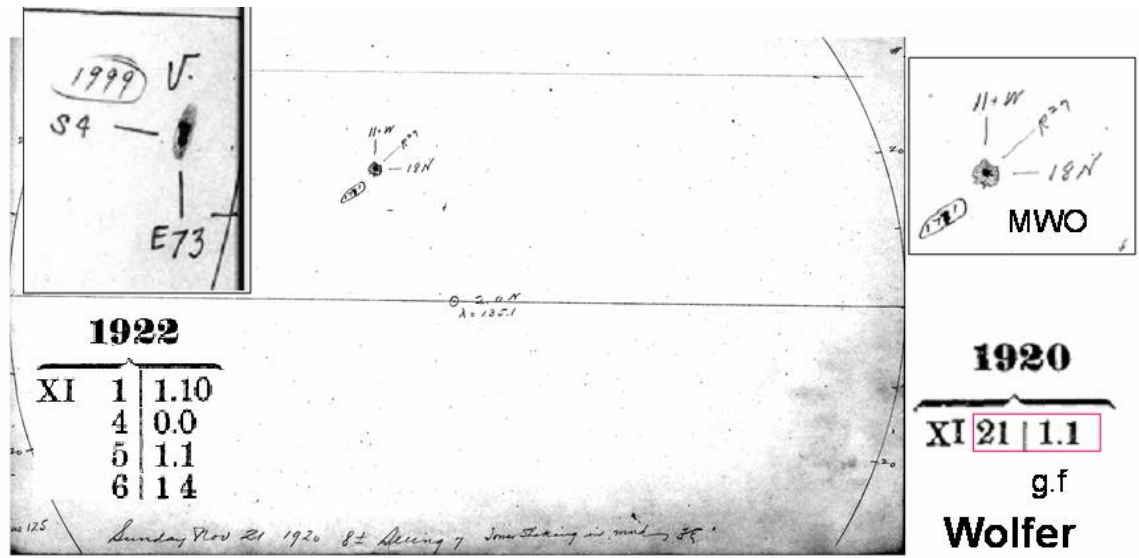

Figure 1: Drawing from Mount Wilson Observatory (MWO) of the single spot with penumbra on 21st Nov. 1920. The insert at the left shows a similar group observed at MWO on 5th Nov., 1922. For both groups, Wolfer should have recorded the observation as "1.3" if he had used the weighting scheme, but they were recorded as "1.1" (one group, one spot), thus counting the large spot only once (*i.e.* with no weighting).

---

[1] A spot like a fine point is counted as one spot; a larger spot, but still without penumbra, gets the statistical weight 2, a smallish spot within a penumbra gets 3, and a larger one gets 5.
[2] When an observer at his instrument on any given day records $g$ groups of spots with a total of $f$ single spots, without regard to their size, then the derived relative sunspot number for that day is $r = k(10g+f)$.

There are many other such examples (*e.g.* 16$^{th}$ September, 1922 and 3$^{rd}$ March, 1924) for which MWO drawings are available at ftp://howard.astro.ucla.edu/pub/obs/drawings and even earlier *e.g.* June 20$^{th}$-23$^{rd}$, 1912 for which we have drawings from the Jesuit-run Haynald Observatory (Kalocsa, Hungary: http://fenyi.sci.klte.hu/deb_obs_en.html, see Slide 11 of http://www.leif.org/research/SSN-workshop1-Weighting.pdf). We can thus consider it established that Wolfer did not apply the weighting scheme. This is consistent with the fact that nowhere in Wolf's and Wolfer's otherwise meticulous yearly reports in the *Mittheilungen über die Sonnenflecken* series is there any mention of a weighting scheme. We remind the reader about the format of Wolf's published observations, Figure 2:

Sonnenfleckenbeobachtungen im Jahre 1849.

| | I. | II. | III. | IV. | V. | VI. | VII. | VIII. | IX. | X. | XI. | XII. |
|---|---|---|---|---|---|---|---|---|---|---|---|---|
| 1 | 9.31 | 3. 6 | 4. – | 10.70 | 9.30 | 8.48 | 4.13 | 4 15 | 7.64 | 8.10 | 5.16 | — |
| 2 | 9.34 | 7.40 | 5. – | 7. – | 9.40 | 9.64 | 3. 3 | 6.18 | 5.35 | 7.10 | 7.41 | 8. 9 |
| 3 | 15. – | 2. – | 6.12 | 10.38 | 5.12 | 8.50 | 3. 6 | 6.15 | 4.27 | 3. 4 | 3.10 | 8.17 |
| 4 | 9.31 | 7.27 | 7.15 | 12.58 | 7.45 | 10.50 | 3 10 | 4.12 | 5.41 | 2. 3 | 4.31 | — |
| 5 | 9. – | 9.22 | 2. – | 8.20 | 8 50 | 8.45 | 7. – | 5.20 | 1. 1 | 1. 2 | — | 9.47 |
| 6 | 8. – | 10 34 | 7.24 | 10.60 | 7.38 | 7.45 | 4. 8 | 4.18 | 6.25 | 4. 6 | — | 2. 2 |
| 7 | — | 3. – | 3. – | 8.24 | 1. – | 5. – | 5.10 | 3.20 | 7.48 | — | 6.22 | — |
| 8 | 8.28 | 10.21 | 4. – | 6.20 | 6.20 | 5.12 | 6.15 | 3.15 | 5.38 | 5.16 | 7.35 | — |
| 9 | 8.30 | 10.35 | 3. – | 9.45 | 6.25 | 3. – | 7.20 | 4.14 | 7.50 | 5.26 | 6.20 | — |

Figure 2: The number of groups *g* and the number of spots (Flecken) *f* for each day is recorded as '*g.f*', (Wolf, 1856). On days where the seeing was poor or when Wolf used a smaller telescope, the entries are in small type font or have no spot count.

To calculate the relative sunspot number, *R*, *e.g.* on April (IV) 4th, Wolf used the well-known formula $R = k \cdot (10 \cdot 12 + 58) = 178$ where the scale factor *k* is 1.00 for Wolf himself.

Clette *et al.* (2014) review the evidence from other solar indices for when the weighting was introduced a well as determining the magnitude of the effect. Svalgaard (2014) provided details of the weighting issue. In the present paper we shall further explore, quantify, and characterize how much the weighting of the sunspot count affects the Relative Sunspot Number.

## 2. Weighting at Locarno: The Reference Station

At the reference station 'Locarno' situated in the city of Locarno on the northern shore of Lago Maggiore in the Swiss canton of Ticino, weighting of the sunspot count has been employed since the beginning in 1957, closely following Waldmeier's prescription (Sergio Cortesi, personal communication). To assess the magnitude of the increase due to weighting, Leif Svalgaard undertook to examine all the nearly 4000 drawings with individual counts of groups and the number of spots in each group made at Locarno (http://www.specola.ch/e/drawings.html) for the past decade (and some years before that) and to re-count the spots without weighting. An example of a drawing with the original weighted counts and the re-counted number of actual spots present is shown in Figure 3. As useful as the drawings are, the final count that is reported to the WDC is that performed visually at the telescope eyepiece and in some cases differs occasionally from the count on the drawing. This is rare enough to not distort the result significantly.

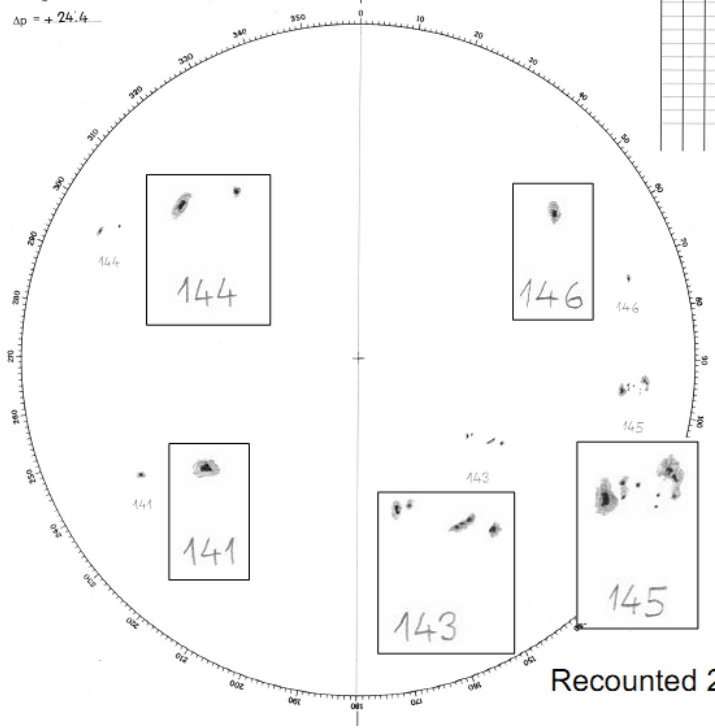

Figure 3: Drawing from Locarno showing the effect of weighting for the five groups present. Magnified views of the groups allow the reader to assess the weighting performed, *e.g.* to see that group 141 consists of one spot with a penumbra, which was assigned weight 3 according to Waldmeier's rule. For this drawing the weight factor of the day becomes 1.36.

At times, the observer did not count the very smallest spots even if they were included in the drawing, Figure 4:

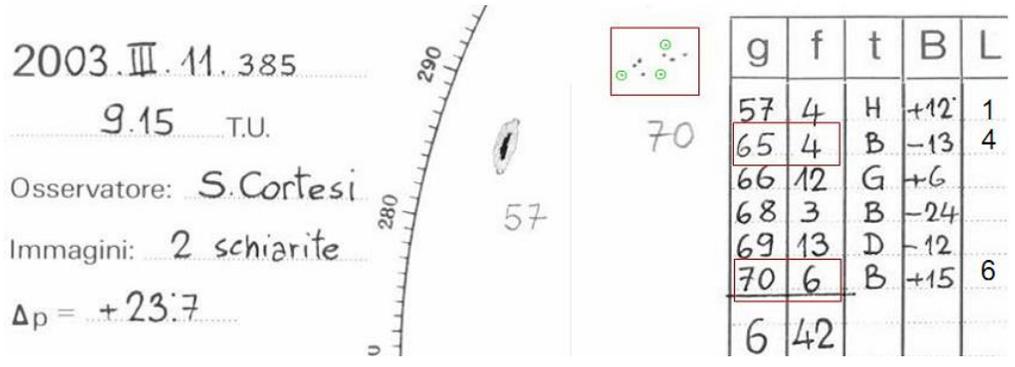

Figure 4: Drawing from Locarno showing tiny spots that were not counted (in green circles) for group number 70. Observers might differ on the 'rule' for omitting tiny spots, but the number of omitted spots is in any case small overall. A useful addition to the report would be the number of omitted spots, if not zero.

In case of the rare very large groups, it is quite a challenge to determine the actual spot count, Figure 5, especially if not all the weakest spots were counted. In this rather extreme case, the top drawing shows 74 spots, but the weighted count is only 58, so clearly many spots (at least 74 – 58 = 16) were not counted. One way to determine the number of un-counted spots would be to weight the large spots (none of which are omitted) according to Waldmeier's prescription, then subtract the sum of all the weighted values, and finally add in the number of spots that were weighted. The Figure shows how that would work. The shaky assumptions underscore the importance of recording the number of omitted spots, or what we could call the 'equivalent' number of omitted spots, if some tiny spots were 'lumped together'.

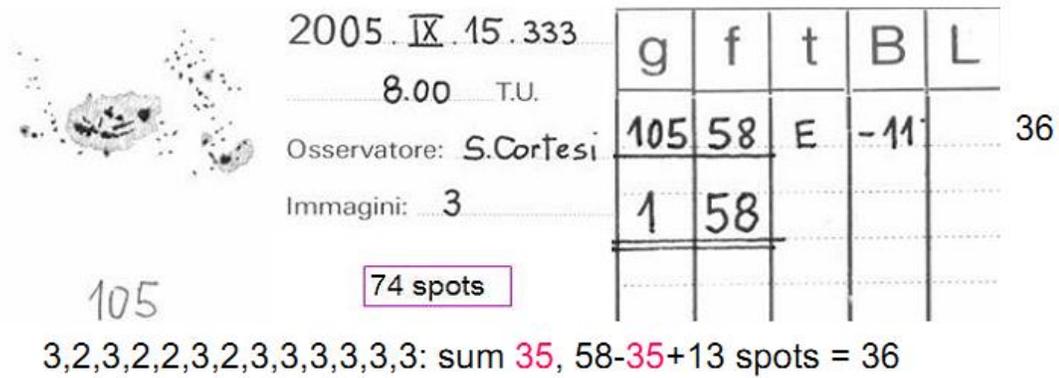

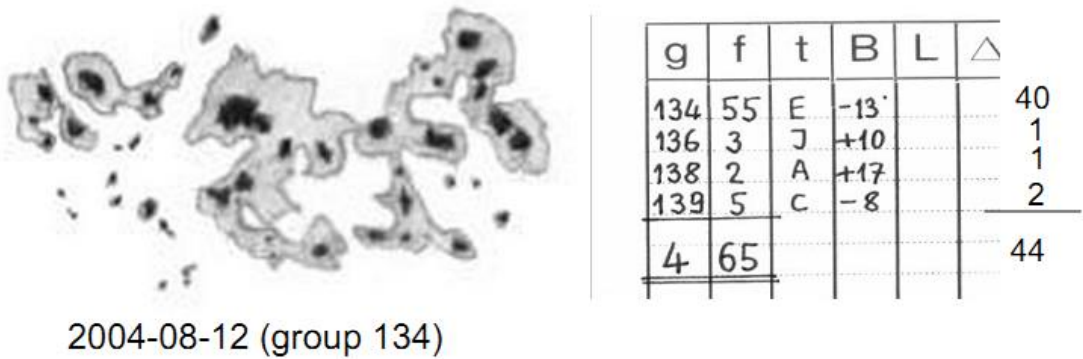

Figure 5: (Top) Drawing from Locarno showing a large, complicated group with many spots that were not counted. The number of spots according to the drawing was 74, but the weighted count was only 58. There were 13 spots (and umbrae) with weights of 3 and 2. The sum of the weighted spots was 35, so the number of spots with weight 1 must be 58 – 35 = 23 to which we must add 13 for a total of (actual?) spots of 36. This example is, admittedly, extreme, but such is the material we have to work with. (Bottom) Drawing of group 134 that on my count had 40 actual spots (and umbrae). The reader is invited to count as well.

To verify that the re-count is valid, i.e. that Svalgaard has understood and applied correctly the Waldmeier weighting scheme, the observer Marco Cagnotti in Locarno had agreed to maintain a (double-blind) parallel count of un-weighted spots at a continuing basis since January 1$^{st}$, 2012, following a brief trial in August 2011, and the un-weighted count is now a part of the routine daily reports. Figure 6 shows that Svalgaard and Cagnotti very closely match each other in applying the weighting scheme, thus sufficiently validating the approach.

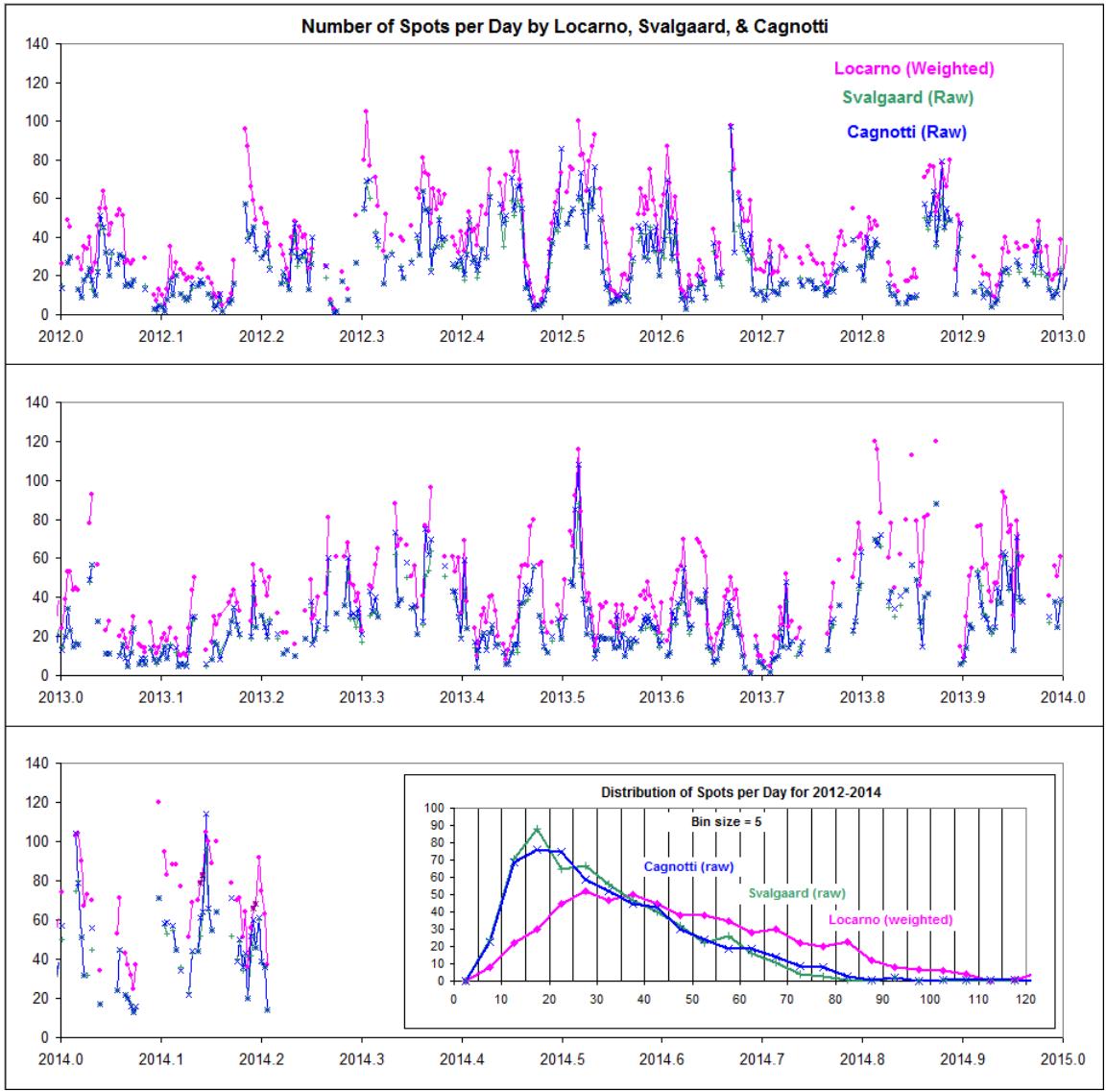

Figure 6: Comparison of the number of sunspots per day determined by Cagnotti (blue) and Svalgaard (green) without weighting, *i.e.* by counting each spot singly as prescribed by Wolfer and Brunner with the number reported by Locarno (pink) employing the Waldmeier weighting scheme. The insert shows the nearly identical distribution of un-weighted counts in bins of five.

Is the weight factor observer dependent? With a novice one might be inclined to think so, but with training, observers tend to converge to agreement. We can compare the weighted counts and the number of groups reported by the veterans Cortesi and Bianda and the new observer Cagnotti from 2008 to the present (Figure 7): there does not seem to be much systematic difference with the possible exception of a very recent decline of Cortesi's weight factor. Observer Andrea Manna (AM) has a weight factor that is systematically about 0.04 lower than the other observers, in spite of seeing the same number of groups, so weighting does depend weakly on the observer.

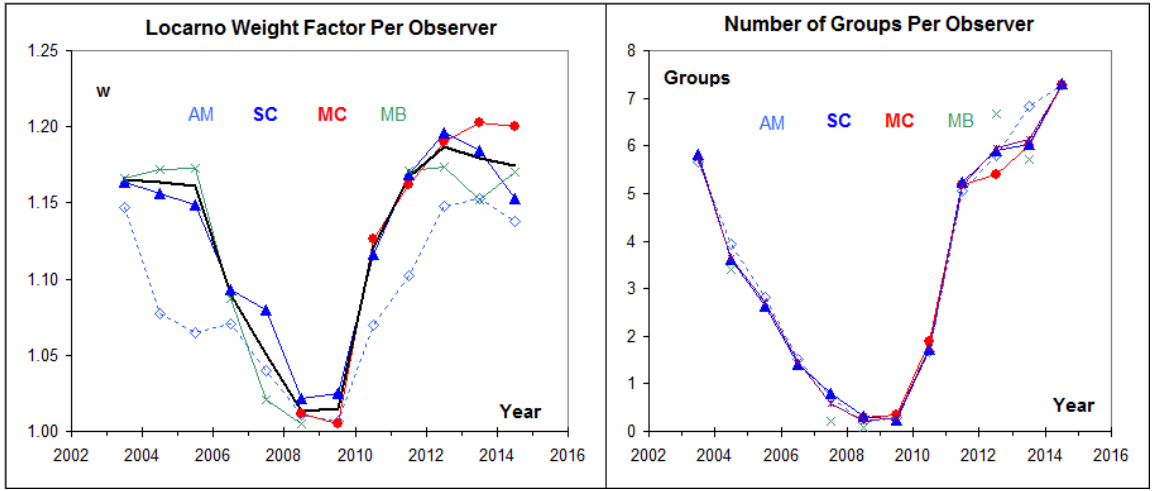

Figure 7: (Left) The weight factor for Locarno observers Cortesi (SC, blue, since 1957), Cagnotti (MC, red, since 2008), Manna (AM, open dashed blue, 1991), and Bianda (MB, green, 1983). (Right) The number of groups per day for each year reported by the same observers.

## 3. The Weighting Quantified by the Locarno Observers

Since August, 2014 the observers in Locarno have augmented their observations of the number of groups, $g$, and of weighted spots, $f$, with a count of actual, non-weighted spots, $s$ (denoted '*LW*' at the right on the drawing – LW is the WDC SIDC/SILSO code designation for un-weighted Locarno counts), allowing us the calculate the weight factor as $w = (10g + f)/(10g + s)$, Figure 8:

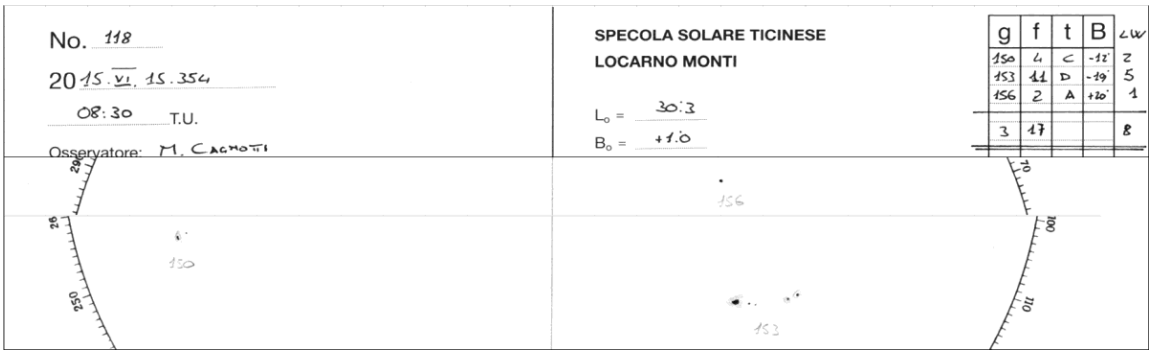

Figure 8: The recent Locarno determination of both the weighted (*f*) and of the un-weighted number of sunspots (*LW*). For this particular day, the weight factor becomes $w = (30+17)/(30+8) = 47/38 = 1.237$.

Figure 9 shows the weight factors determined from the Locarno observations since August, 2014. The red curve shows the 27-day running average of the weight factor calculated using the relationship determined by Clette *et al.* (2014). It is clear that the Clette *et al.* (2014) expression for the weight factor agrees well with the observations for this level of solar activity. It is also clear that the value (1.116) marked by the blue line, as was suggested by Lockwood *et al.* (2014), is not a good fit to the observations and as such must be discarded.

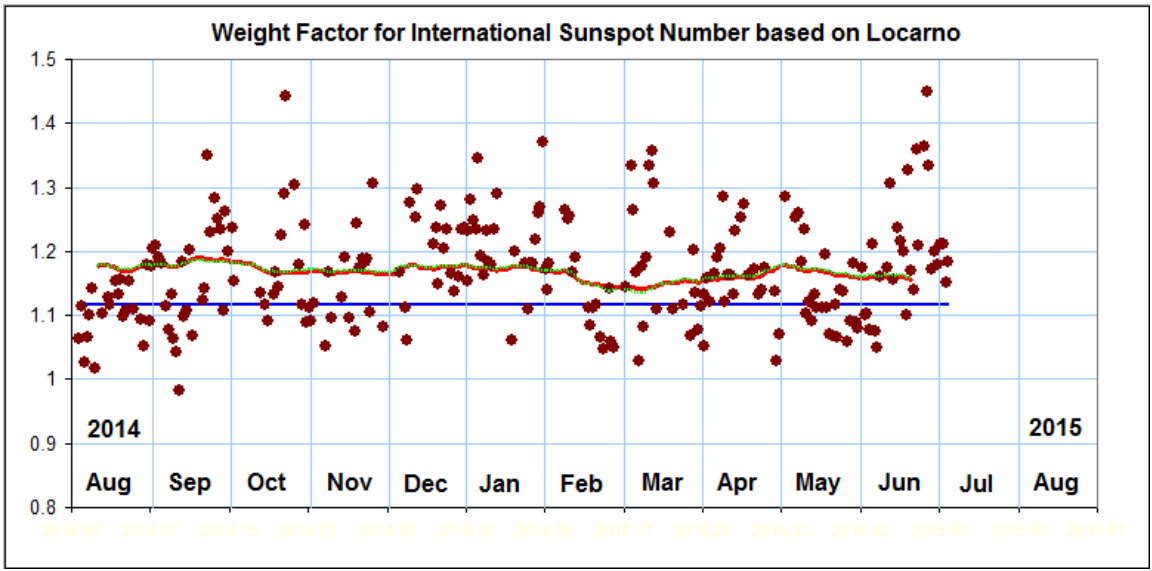

Figure 9: Weight factors (brown dots) computed from the recent Locarno daily data. The red curve shows the 27-day running average of the weight factor calculated using the relationship determined by Clette *et al.* (2014).

Figure 10 shows the Locarno weight factor as determined by Svalgaard (blue symbols) for both solar maximum and solar minimum conditions and continued (red symbols) by the Locarno observers until the present [and hopefully beyond]. The green dots show yearly averages.

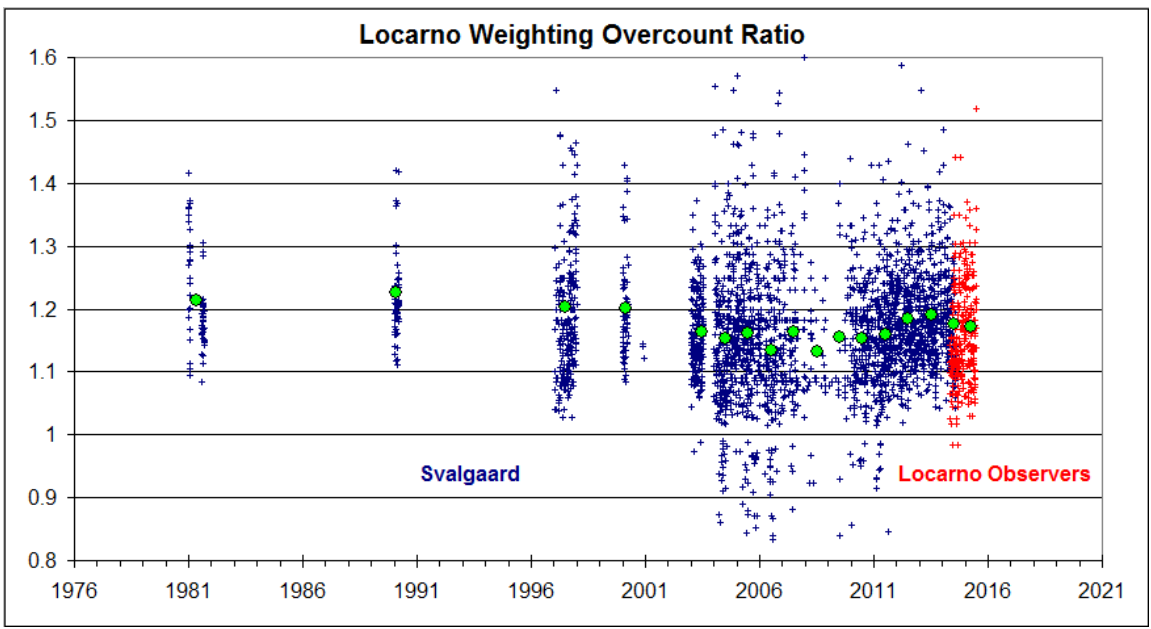

Figure 10: Locarno weight factor as determined by Svalgaard (blue symbols) for both solar maximum and solar minimum conditions and continued (red symbols)

by the Locarno observers until the present. The green dots show yearly averages and a weak solar cycle modulation.

The problem we are faced with is not really to calculate the weight factor for the current data. We don't need to; we know what the factor is for every day (with an observation). The problem is to determine the weight factor retroactively for the interval 1947-1980. The Locarno drawings 1957-1980 and the Zürich archives 1937-1980 are apparently lost (thrown out? Or to use a modern term 'have been disappeared'). For the Zürich data before 1980 we know the number of groups for each month and the relative sunspot number (encumbered by weighting because all observers were normalized to Zürich) for each day (and hence for each month). Can we from that correct the sunspot number for weighting? Before we attack that problem, we'll look closer at the data on a daily basis.

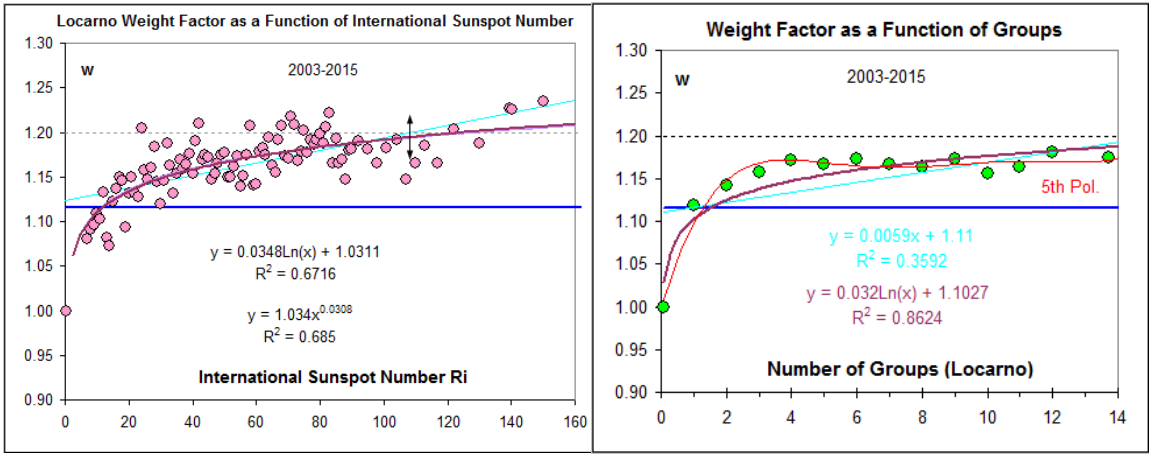

Figure 11: (Left) Locarno average weight factor for bins of the International Sunspot Number (*Ri*). Below *Ri* = 90, the bin size is unity, while above that, bins of progressively larger size are used to ensure enough values in each bin. A fiducial value of 0.3 has been used in lieu of a zero *Ri* to which a weight of 1 is assigned. The double-headed arrow shows an estimate of the error of the mean values. (Right) Average weight factor for unity-wide bins of the number of groups. A fiducial value of 0.1 has been used in lieu of a zero group number, to which a weight of 1 has been assigned.

On a daily basis, the dependences of the weight factor on *Ri* and on the number of groups are decidedly non-linear with a rapid drop-off towards low activity, but even a slightly wrong weight factor applied to a small value will have very little effect on the result. But it is clear that the daily weight factor is not just a simple function of the relative number *SSN* or of the group count alone, *GN*, but is a function of both (and of the observer as well): *w* = *F* (*SSN*, *GN*, *Obs*). The situation is further complicated by *SSN* being also a function of *GN*, *Obs*, and of the number of spots, *SN*: *SSN* = *Q* (*GN*, *SN*, *Obs*), so that we actually should write *w* = *F*(*Q*(*GN*, *SN*, *Obs*),*Obs*). As the dependence on the Zürich observers is slight, we ignore the observer differences as furthermore also necessitated by the fact that we don't know who the observers were for each day during 1947-1980. To separate the influence of *GN* and *SN* we now plot the daily Locarno weight factor as a function of the reported (*i.e.* weighted) *SN* for bins of each group number, Figure 12:

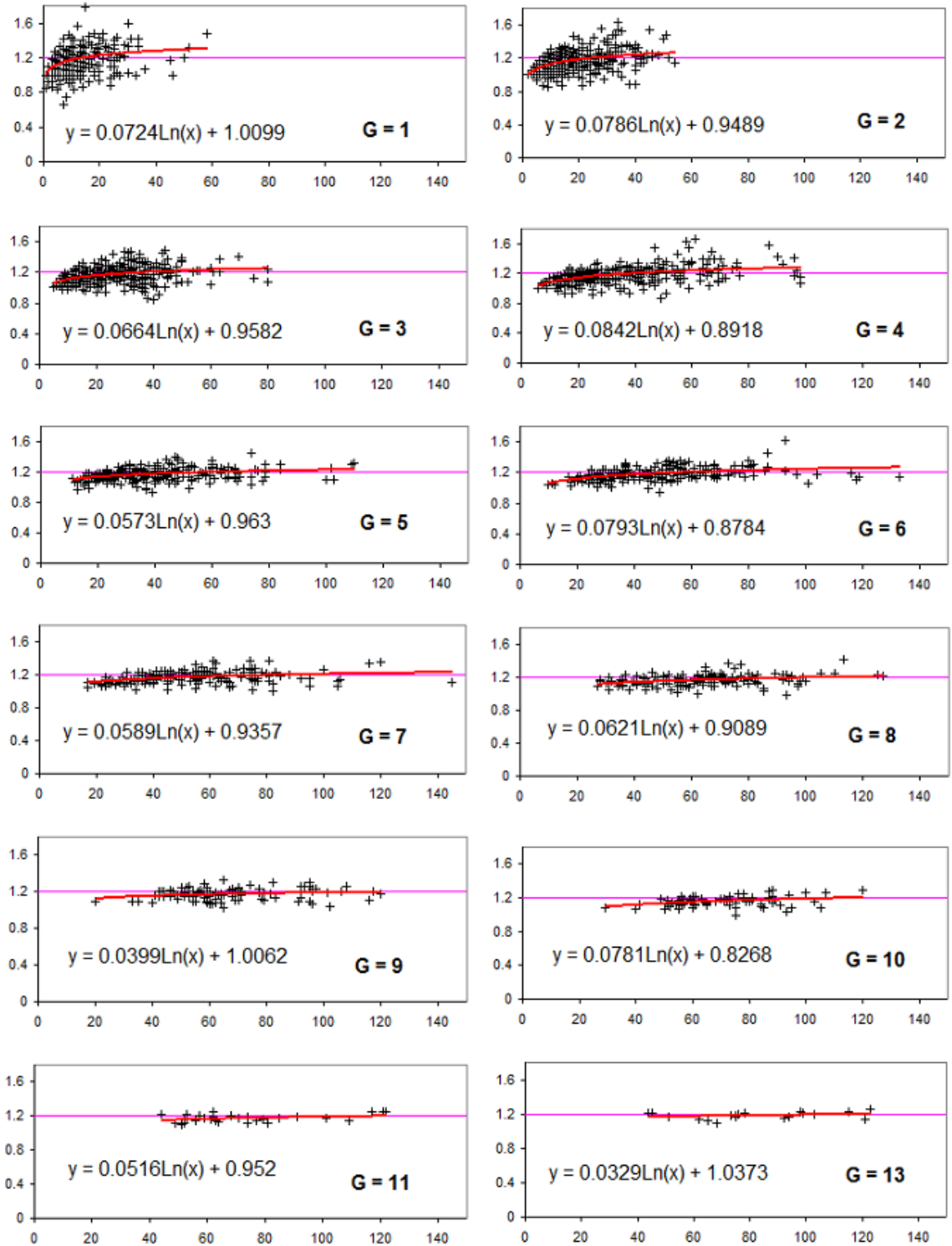

Figure 12: For each bin of group number G = 1, 2, 3, … the graphs show the Locarno weight factor for re-counted days of 1997-2015 as a function of the reported (thus weighted) number of sunspots (note: *not* the sunspot number). A fit to a logarithmic function of the sunspot count is derived for each group.

Using the functional fits derived from Figure 12 we calculate the weight factor on a grid of 1 unit of GN and 5 units of SN to obtain a visual representation of the weight factor 'landscape' function $w = F(Q(GN, SN))$, Figure 13 (left panel). The 'jagged' appearance could be improved by suitable smoothing, but the gain seems marginal. We can thus quantify the average effect of Weighting given the group and (reported) spot counts for daily values, should such values become available.

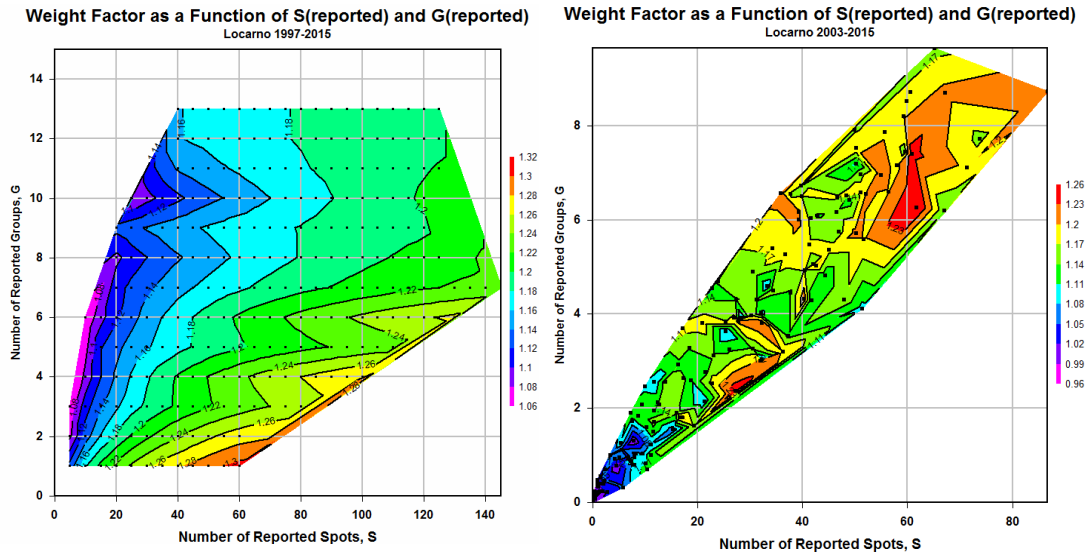

Figure 13: (Left) Contour map of the daily Locarno weight factor for 1997-2015 as a function jointly of the reported (thus weighted) number of sunspots, *S*, and of the number of groups, *G*. (Right) Contour map of the monthly Locarno weight factor for 2003-2015 as a function of both *S* and *G*.

It is also of interest to repeat the analysis for monthly values, *e.g.* as given in Waldmeier (1968b, 1978), as the scatter is much smaller, *c.f.* Figure 13 (right). The results are shown in Figure 14 and 13 (right panel).

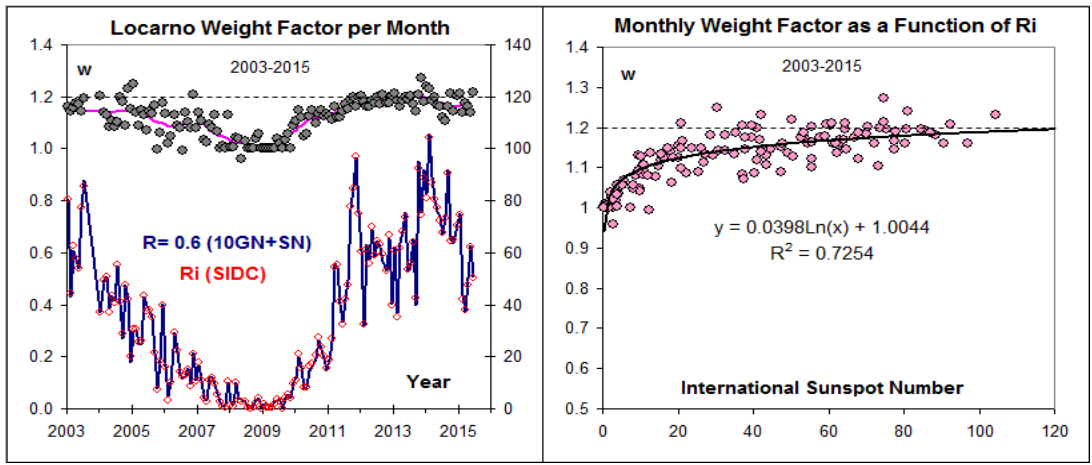

Figure 14: (Left) The Locarno weight factor for each month for 2003-2015 dipping down to unity for no activity and rising to 1.2 for the moderate activity

at the maximum of the weak solar cycle 24. At the bottom we show the time variation of the International Sunspot Number (red circles) which is very closely the same as the Locarno relative number multiplied by the nominal *k*-factor of 0.60 (blue curve). (Right) The monthly weight factors as a function of the International Sunspot Number. The non-linear function shown is a decent fit to the weight factor data.

## 4. Correcting for Weighting

For monthly values, the group count and the spot count are constrained to a rather narrow diagonal band in Figure 13 (right) which suggests that a one-dimensional relationship with the relative sunspot number, such as given in Figure 14 (right), might be sufficient for correction of said number to an un-weighted value. We can test this assertion by calculating the weight factor using that formula ($w = 1.0044 + 0.0398 \ln(R_i)$; $R_i \geq 0.2$), dividing the International Sunspot Number since 2003 by the computed weight factor, and comparing the thus corrected number with the un-weighted relative number obtained by re-counting the spots without weighting on the Locarno drawings, Figure 15. The agreement is excellent, with a linear coefficient of determination $R^2 = 0.991$:

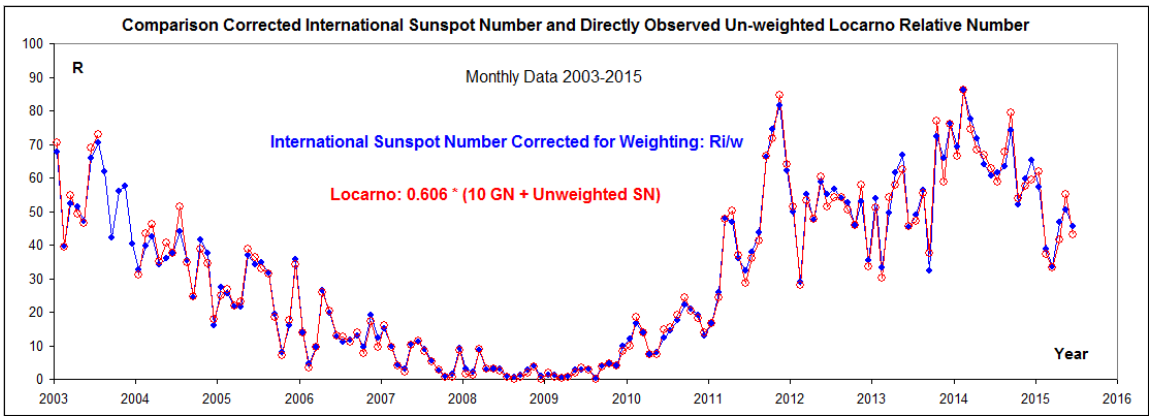

Figure 15:.Comparison of monthly values of the International Sunspot Number as published by the WDC SILSO in Brussels (Version 1.0, pre-July-1$^{st}$-2015) corrected for weighting (blue curve) and the Relative Number for Locarno calculated using the un-weighted number of sunspots (red curve) and a *k*-factor of 0.606.

Under the assumption that the weight factor function is also valid for the Waldmeier era at Zürich we can now correct the Zürich sunspot number for the inflation introduced by the weighting scheme, Figure 16 and Table 1.

In constructing Figure 16 (and in this paper generally) we used the pre-July-1$^{st}$-2015 values of the International Sunspot Number without the corrections and reassessments introduced as of that date. It is important to take into account that the weight factor varies with the sunspot number itself, so one cannot (except as a first, crude approximation) use a constant weight factor throughout. The average yearly weight factors given in Table 1 are valid regardless of the sunspot numbers determined for each year and of the *k*-factors adopted. The factors were derived from the formula of Figure 14 using the nominal *k*-factor of 0.60, so its $R_i$-argument could be written $R_i = R_k*0.6/k$, where *k* is the *k*-factor for the relative sunspot number $R_k$. For $R_k$ from the 'new' SILSO sunspot number series, $k$ is equal to unity.

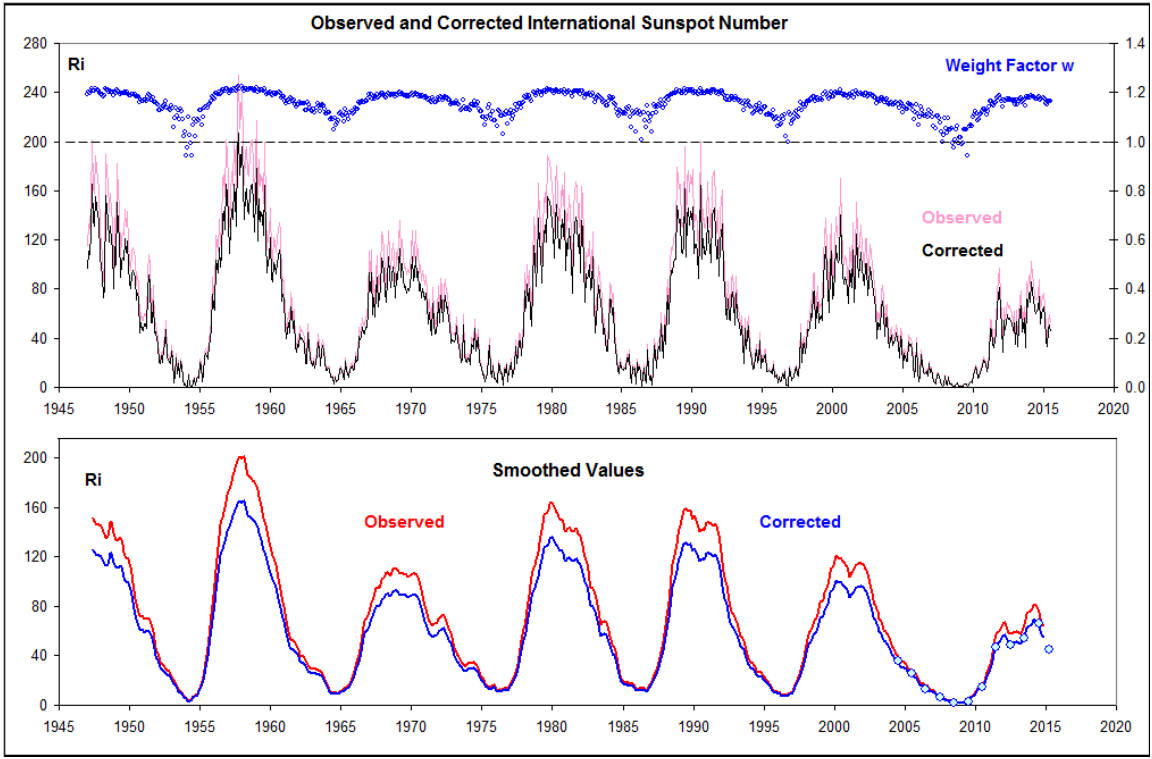

Figure 16: (Top) Comparison of monthly values of the International Sunspot Number as published by the WDC SILSO in Brussels (Version 1, pre-July-1$^{st}$-2015), pink curve, and the values corrected for weighting (black curve) using the weight factors shown by the upper blue symbols. (Bottom) The monthly values smoothed (using the standard method introduced by Wolf). Light blue dots show yearly values of un-weighted counts from Locarno, *i.e.* not relying on the weight factor formula. Again, the agreement is excellent.

An interesting question is: how does this 'corrected New $R_i$' (which is simply SILSO V1 $R_i$ freed from weighting and brought onto Wolfer's scale by removing the obsolete 0.6 $k$-value scale factor, call it V1.5) compare with WDC-SILSO V2 $R_i$ released July 1$^{st}$, 2015? Figure 17 provides a preliminary answer to that question:

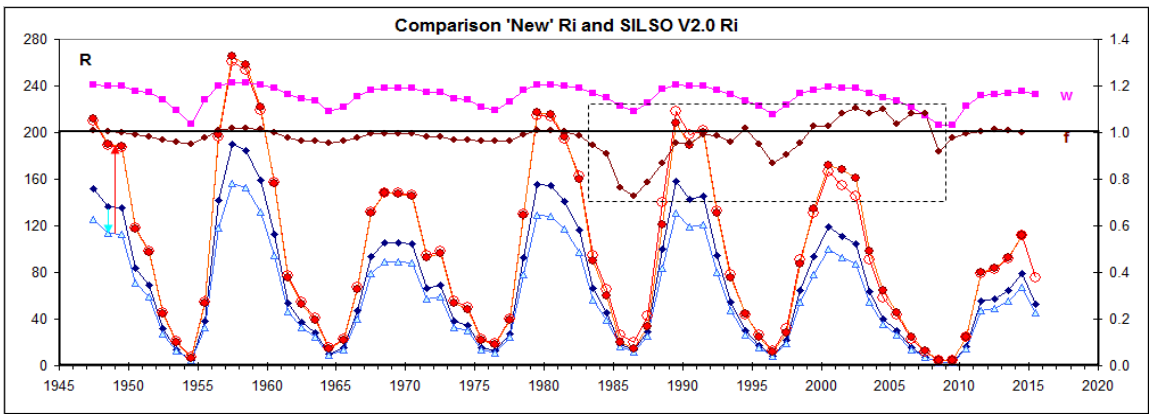

Figure 17: Dark blue diamonds (V1 Ri - old official Ri], scaled down to the 'Corr. Ri', light blue triangles [V1.5], by dividing by the weight factor, $w$, (upper pink squares). The 'Corr. Ri' is then scaled to the Wolfer scale (New Ri, red open circles) by dividing by the, no longer used $k$-value 0.60 and compared with SILSO V2 Ri (red filled circles). The ratio $f$ = V2/New is shown by the brown dots.

The ratio $f$ = V2/New (brown dots) is generally close to unity, although there is a weak solar cycle variation, probably due to an inadequate $w$-factor used for SILSO V2. The ratio varies irregularly for the years in the rectangle, possibly indicating some further (unexplained) adjustments. The irregularity is not serious near solar minima, as the sunspot number is small then, but the ~10% difference at the maximum and declining part of sunspot cycle 23 is a concern that should be addressed and explained.

## 5. Conclusions

Waldmeier in 1947 introduced a weighting (on a scale from 1 to 5) of the sunspot count made at Zurich and its auxiliary station Locarno, whereby larger spots were counted more than once. This counting method inflates the relative sunspot number over that which corresponds to the scale set by Wolfer and Brunner. Svalgaard re-counted some 60,000 sunspots on drawings from the reference station Locarno and determined that the number of sunspots reported were 'over counted' by 44% on average, leading to an inflation (measured by a weight factor) in excess of 1.2 for high solar activity. In a double-blind parallel counting by the Locarno observer Cagnotti, we determined that Svalgaard's count closely matches that of Cagnotti's, allowing us to determine the daily weight factor since 2003 (and sporadically before). We find that a simple empirical equation fits the observed weight factors well, and use that fit to estimate the weight factor for each month back to the introduction of weighting in 1947 and thus to be able to correct for the over-count and to reduce sunspot counting to the Wolfer method in use from 1893 onwards. The Locarno observers have since August, 2014 counted spots both with and without weighting, and the un-weighted (real) spot count is now used in determining the official relative sunspot number.

================================================================

Table 1: `Old` $R_i$ is the International Sunspot Number (version 1.0), `Corr.` $R_i$ is `Old` $R_i$ divided by the Weight Factor (calculation actually performed month-by-month, then averaged per year). 'New' $R_i$ is `Corr.` $R_i$ divided by 0.60, but does not quite match SILSO version 2.0 because of further (small) corrections to the latter.

| The Year | Old $R_i$ | Weight Factor | Corr. $R_i$ | 'New' $R_i$ | | The Year | Old $R_i$ | Weight Factor | Corr. $R_i$ | 'New' $R_i$ |
|---|---|---|---|---|---|---|---|---|---|---|
| 1947.5 | 151.5 | 1.204 | 125.8 | 209.7 | | 1982.5 | 116.3 | 1.193 | 97.4 | 162.3 |
| 1948.5 | 136.2 | 1.199 | 113.4 | 189.0 | | 1983.5 | 66.6 | 1.169 | 56.8 | 94.7 |
| 1949.5 | 135.1 | 1.199 | 112.6 | 187.7 | | 1984.5 | 45.9 | 1.149 | 39.4 | 65.7 |
| 1950.5 | 83.9 | 1.179 | 71.0 | 118.3 | | 1985.5 | 17.9 | 1.115 | 16.0 | 26.7 |
| 1951.5 | 69.4 | 1.172 | 59.1 | 98.5 | | 1986.5 | 13.4 | 1.093 | 12.0 | 20.0 |
| 1952.5 | 31.4 | 1.140 | 27.5 | 45.8 | | 1987.5 | 29.2 | 1.129 | 25.5 | 42.5 |
| 1953.5 | 13.9 | 1.096 | 12.4 | 20.7 | | 1988.5 | 100.0 | 1.185 | 84.0 | 140.0 |
| 1954.5 | 4.4 | 1.035 | 4.1 | 6.8 | | 1989.5 | 157.8 | 1.205 | 130.8 | 218.0 |

| Year | | | | | Year | | | | |
|---|---|---|---|---|---|---|---|---|---|
| 1955.5 | 38.0 | 1.139 | 32.8 | 54.7 | 1990.5 | 142.3 | 1.201 | 118.9 | 198.2 |
| 1956.5 | 141.7 | 1.200 | 117.8 | 196.3 | 1991.5 | 145.8 | 1.202 | 121.2 | 202.0 |
| 1957.5 | 189.9 | 1.212 | 156.4 | 260.7 | 1992.5 | 94.5 | 1.184 | 79.6 | 132.7 |
| 1958.5 | 184.6 | 1.212 | 152.3 | 253.8 | 1993.5 | 54.7 | 1.162 | 47.0 | 78.3 |
| 1959.5 | 158.8 | 1.205 | 131.5 | 219.2 | 1994.5 | 29.9 | 1.137 | 26.1 | 43.5 |
| 1960.5 | 112.3 | 1.192 | 94.1 | 156.8 | 1995.5 | 17.5 | 1.115 | 15.6 | 26.0 |
| 1961.5 | 53.9 | 1.162 | 46.3 | 77.2 | 1996.5 | 8.6 | 1.079 | 7.9 | 13.2 |
| 1962.5 | 37.6 | 1.147 | 32.7 | 54.5 | 1997.5 | 21.5 | 1.118 | 18.9 | 31.5 |
| 1963.5 | 27.9 | 1.135 | 24.5 | 40.8 | 1998.5 | 64.2 | 1.168 | 54.8 | 91.3 |
| 1964.5 | 10.2 | 1.092 | 9.3 | 15.5 | 1999.5 | 93.2 | 1.183 | 78.5 | 130.8 |
| 1965.5 | 15.1 | 1.110 | 13.5 | 22.5 | 2000.5 | 119.5 | 1.194 | 100.0 | 166.7 |
| 1966.5 | 46.9 | 1.156 | 40.4 | 67.3 | 2001.5 | 110.9 | 1.191 | 93.0 | 155.0 |
| 1967.5 | 93.7 | 1.184 | 79.0 | 131.7 | 2002.5 | 104.1 | 1.189 | 87.5 | 145.8 |
| 1968.5 | 105.9 | 1.190 | 89.0 | 148.3 | 2003.5 | 63.6 | 1.169 | 54.3 | 90.5 |
| 1969.5 | 105.6 | 1.190 | 88.7 | 147.8 | 2004.5 | 40.4 | 1.150 | 35.1 | 58.5 |
| 1970.5 | 104.7 | 1.189 | 88.0 | 146.7 | 2005.5 | 29.8 | 1.136 | 26.1 | 43.5 |
| 1971.5 | 66.7 | 1.171 | 56.9 | 94.8 | 2006.5 | 15.2 | 1.109 | 13.5 | 22.5 |
| 1972.5 | 68.9 | 1.172 | 58.7 | 97.8 | 2007.5 | 7.5 | 1.073 | 6.9 | 11.5 |
| 1973.5 | 38.2 | 1.147 | 33.1 | 55.2 | 2008.5 | 2.9 | 1.034 | 2.7 | 4.5 |
| 1974.5 | 34.4 | 1.143 | 30.0 | 50.0 | 2009.5 | 3.1 | 1.033 | 2.9 | 4.8 |
| 1975.5 | 15.5 | 1.107 | 13.8 | 23.0 | 2010.5 | 16.5 | 1.114 | 14.8 | 24.7 |
| 1976.5 | 12.6 | 1.097 | 11.3 | 18.8 | 2011.5 | 55.6 | 1.161 | 47.6 | 79.3 |
| 1977.5 | 27.5 | 1.132 | 24.1 | 40.2 | 2012.5 | 57.6 | 1.165 | 49.4 | 82.3 |
| 1978.5 | 92.7 | 1.183 | 78.1 | 130.2 | 2013.5 | 64.7 | 1.169 | 55.2 | 92.0 |
| 1979.5 | 155.3 | 1.205 | 128.8 | 214.7 | 2014.5 | 79.1 | 1.178 | 67.1 | 111.8 |
| 1980.5 | 154.7 | 1.205 | 128.3 | 213.8 | *2015.3* | *52.7* | *1.162* | *45.4* | *75.7* |
| 1981.5 | 140.5 | 1.201 | 116.9 | 194.8 | | | | | |

## Acknowledgements

We have benefited from the Sunspot Number Workshops and from discussions with the team at the WDC/SILSO. LS thanks Stanford University for support.